\documentclass[12pt]{iopart}
\usepackage[utf8]{inputenc}
\usepackage{graphicx}
\usepackage{epstopdf}
\usepackage{color}
\usepackage[dvipsnames]{xcolor}
\usepackage{siunitx}
\usepackage[outercaption]{sidecap}
\usepackage{rotating}
\usepackage{pdflscape}
\usepackage{comment}
\usepackage{array}
\usepackage{adjustbox}
\usepackage{gensymb}

\begin{document}

\title[High yield creation of germanium vacancy centers in diamond]{High yield creation of germanium vacancy centers in diamond by focused ion beam implantation and high temperature annealing}

\author{{S Dietel}$^{1,\dagger}$, {E Scattolo}$^{2,\dagger}$, {J Fuhrmann}$^1$, {L Kazak}$^1$, A Cian$^2$, E Missale$^2$, {D Giubertoni}$^2$ and F Jelezko$^{1,3}$}
\address{$^1$ Institute for Quantum Optics, Ulm University, Albert-Einstein-Allee 11, 89081 Ulm, DE}
\address{$^2$ Center for Sensors and Devices, Fondazione Bruno Kessler, Via Sommarive 18, I-38123 Trento, IT}
\address{$^3$ Center for Integrated Quantum Science and Technology (IQST), Albert-Einstein-Allee 11, 89081 Ulm, DE}
\ead{stefan.dietel@uni-ulm.de}

\begin{indented}
\item[] $^\dagger$These authors have contributed equally
\end{indented}

\begin{abstract}
Negatively charged germanium vacancy centers (GeV) in diamond are a promising platform for quantum computing and quantum communication. However, these applications require the precise incorporation of GeV centers with good optical properties inside of nanophotonic structures. In this work, we demonstrate the highly efficient local creation of GeV centers in diamond via focused-ion-beam implantation, followed by high-temperature annealing. We report the successful creation of GeV centers over the depth range of 5.5 - 30\,nm. Implantation at low fluence enables the creation of single GeV centers. The formation yield strongly depends on implantation energy and fluence, reaching up to 33\,\% at energies of 35 and 70\,keV. This method, therefore, enables the efficient creation of GeV centers within a small, well-defined local sample volume and offers a potential means of incorporating them into photonic structures.

\end{abstract}


\maketitle

\section{Introduction}

Diamond is an appealing material for quantum technologies due to its large energy gap, high hardness, high thermal conductivity, and chemical stability.
The wide band gap of 5.5\,eV enables diamond to host numerous optically active point defects, known as color centers (CCs), which are associated with deep electronic levels \cite{zhang2020material}. 
These CCs exhibit atom-like properties, such as electronic spin states, which can be exploited for diverse applications, even under ambient conditions \cite{awschalom2018,bradac2019}.
The most prominent and best-studied CC is the nitrogen-vacancy (NV) center \cite{Bala2009NV, Doherty2013NV}. The NV center enables precise spin manipulation and readout at both room and cryogenic temperatures, demonstrating promising results in applications such as magnetic field sensing \cite{Rondin2014NVMagn} and temperature sensing \cite{Neumann2013Temp, Kucsko2013Temp}.
For advanced applications, efficient collection of emitted light and precise control of light-matter interaction are necessary. This can be achieved by integrating NV centers into nano-photonic structures. However, this is due to the weak emission into the zero-phonon line and the strong sensitivity of NV centers to local strain \cite{Maze2011NVstrain}. Additionally, the close proximity to the diamond surface can induce spectral \cite{Chakravarathi_2021} and charge instability \cite{Fuhrmann2024NVsurface}.

A promising alternative is the group IV color (G4V) centers, such as the 
silicon-, germanium-, tin-, and lead-vacancy centers \cite{Neu_2011,Iwasaki2015-op,Iwasaki_SnV2017,Ditalia_Tchernij2018}. The G4V center belongs to the $D_{3d}$ symmetry group and therefore has an inversion symmetry, leading to the absence of a permanent electric dipole moment. It results in low sensitivity to electric fields (only in second order), local strain, charge-related noise, and decoherence. These characteristics make them particularly well-suited for applications typically limited by environmental perturbations. Moreover, G4V centers exhibit strong zero-phonon line emission together with long electron \cite{Sukachev_SiV2017,Karapatzakis_SnV2024} and nuclear spin coherence times \cite{Metsch_SiV2019,Resch_SnV2026}. In particular, the negatively charged germanium-vacancy (GeV) color center shows a record electron spin coherence time of up to 20\,ms \cite{SenPRL24} and nuclear spin coherence times of up to 2.5\,s \cite{GrimmPRL25}, making it a promising candidate for quantum communication applications, once integrated into nano-photonic structures such as photonic crystal cavities, pillars, or waveguides \cite{Pugliese2025-le, Redolfi2025-ju}.

\newcolumntype{C}[1]{>{\centering\arraybackslash}p{#1}}
\newcolumntype{L}[1]{>{\raggedright\arraybackslash}p{#1}}

\begin{table}[p]
\centering
\caption{Summary of the fabrication processes reported in literature to produce GeV centers in diamond, together with the achieved main optical properties.}
\label{table:summary}

\fontsize{12}{11}\selectfont   
\renewcommand{\arraystretch}{1.3}
\setlength{\extrarowheight}{3pt}
\setlength{\tabcolsep}{10pt}

\begin{adjustbox}{max width=\textwidth}
\begin{tabular}{C{0.8cm} C{0.8cm} L{2.4cm} L{2.6cm} C{2.2cm} L{2.8cm} C{1.4cm} C{1.6cm} C{1.4cm} C{1.2cm}}
\br
Ref. & Year & Substrate \newline structure& Implantation energy \newline (keV) & Ion fluence \newline range \newline \si{\mathrm{(ions/cm^2)}} & Annealing T,\newline time, atm. & Formation\newline yield & SPE ion \newline fluence \newline \si{\mathrm{(ions/cm^2)}} & $I_{sat}$\newline (kcps) & $P_{sat}$\newline (mW)\\
\mr

\cite{Iwasaki2015-op}&2015&Electronic grade\newline bulk& 150-260 & $3.5 \times 10^{8}$  \newline $5.9 \times 10^{13}$ &$\SI{800}{\celsius}$,\newline 3h&-& $3.5 \times 10^{8}$&75-170&-\\

\cite{Bhaskar2017-ig}&2017& Tapered waveguide& 275 &$1 \times 10^{9}$ & $\SI{1200}{\celsius}$ &-&$1 \times 10^{9}$& WG:\newline $790 \pm 20$ \newline bulk:\newline $560 \pm 20$& $4.7 \pm 0.2$\\

\cite{ZhouNJP18}&2018&[N]$<$1 ppb\newline bulk&FIB\newline 70& 100, 200, 400, 700 i+&$\SI{1000}{\celsius}$,\newline 30 min, high vacuum&0.60$\%$&100-200 i+&$178 \pm 4$&$4.5 \pm 0.2$\\

\cite{Wan2020}&2020&single-crystal diamond plate& 200 & $2-6 \times 10^{11}$ & $\SI{1200}{\celsius}$, $1 \times 10^{-9}$ mbar& $1.9\%$&-&$640 \pm 360$&-\\

\cite{Chen2022-di}&2022&IIa-type\newline SIL& 10$^4$ & $1 \times 10^{11}$ & $\SI{100}{\celsius}$ in 2h,\newline hold for 11h,\newline $\SI{400}{\celsius}$ in 4h,\newline hold for 8h,\newline $\SI{800}{\celsius}$ in 12h,\newline hold for 8h,\newline $\SI{1100}{\celsius}$ in 12h,\newline hold for 8h&-&-&$\sim$16&$0.183 \pm 0.02$\\

\cite{Nieto_Hernandez2023-no}&2023&Electronic grade\newline bulk& 40&$1 \times 10^{10}$-$1 \times 10^{13}$ &$\SI{900}{\celsius}$, 2h,\newline $<5 \times 10^{-6}$& 0.60$\%$&$3 \times 10^{12}$&$900 \pm 600$& $3.0 \pm 1.2$\\

\cite{Nieto_Hernandez2023-no}&2023&Electronic grade\newline bulk&40&$1 \times 10^{10}$-$1 \times 10^{13}$ &$\SI{900}{\celsius}$, 2h, $<5 \times 10^{-6}$ mbar \newline $\SI{1000}{\celsius}$, 10h, $<5 \times 10^{-6}$ mbar& 0.60$\%$&$3 \times 10^{12}$ &$600 \pm 500$& $3.1 \pm 1.4$\\

\cite{Nieto_Hernandez2023-no}&2023&Electronic grade\newline bulk&40&$1 \times 10^{10}$-$1 \times 10^{13}$ &$\SI{1500}{\celsius}$, 1h,\newline $<5 \times 10^{-6}$ mbar& 0.60$\%$&$3 \times 10^{12}$ &$650 \pm 350$& $1.2 \pm 0.5$\\

\cite{Nieto_Hernandez2023-no}&2023&Optical grade\newline bulk&40&$1 \times 10^{10}$-$1 \times 10^{13}$ &Growth: $\SI{2000}{\celsius}$,\newline 15 min, 6 GPa\newline $\SI{950}{\celsius}$,\newline $<5 \times 10^{-6}$ mbar& 0.60$\%$&$3 \times 10^{12}$ &$1200 \pm 400$& $0.9 \pm 0.3$\\

\cite{Christinck2023-od}&2023&Detector grade\newline SIL&3000& $1 \times 10^{11}$ & $\SI{1200}{\celsius}$, 2h,\newline $<5 \times 10^{-6}$ mbar&-&$1 \times 10^{11}$ & $854 \pm 8$&  $5.7 \pm 0.1$\\

\cite{Hoy_Jensen2020-xv}&2024&Electronic grade\newline $\SI{1}{\mu m}$ thick membrane&330& $1 \times 10^{9}$ &$\SI{400}{\celsius}$ in 1h,\newline hold for 4,\newline $\SI{800}{\celsius}$ in 1h,\newline hold for 2h,\newline $\SI{1200}{\celsius}$ in 1h,\newline hold for 2h,\newline $<1 \times 10^{-6}$ Torr&6$\%$&$1 \times 10^{9}$ &$52 \pm 7$&-\\

\cite{WahlMQT24}&2024&Electronic grade\newline bulk&30 (at $\SI{17}{\degree}$)\newline Beta decay recoil& $2 \times 10^{12}$\newline $2 \times 10^{13}$  & $\SI{300}{\celsius}$, $\SI{600}{\celsius}$, $\SI{900}{\celsius}$,\newline 10 min,\newline $<1 \times 10^{5}$ mbar&-&-&-&-\\

\cite{Zifkin2024-id}&2024&Electronic grade\newline $\SI{1}{\mu m}$ thick membrane&330&$1 \times 10^{9}$ & $\SI{400}{\celsius}$ in 1h,\newline hold for 4,\newline $\SI{800}{\celsius}$ in 1h,\newline hold for 2h,\newline $\SI{1200}{\celsius}$ in 1h,\newline hold for 2h &-&$1 \times 10^{9}$ &-&$52 \pm 7$\\

\cite{Redolfi2025-ju}&2025&Electronic grade\newline Nanopillars on $\SI{50}{\mu m}$ memb. &FIB\newline 70, 35& 30 - 800 \si{\mathrm{ions/spot}}& $\SI{1000}{\celsius}$,\newline 2h, $<1 \times 10^{-6}$ Torr&-&35 \si{\mathrm{ions/spot}} at 35 keV&$268 \pm 15$& $7.6 \pm 0.6$\\

\cite{Pugliese2025-le}&2025&Electronic grade\newline bulk&FIB\newline 70& 7-660\newline \si{\mathrm{ions/spot}}& $\SI{900}{\celsius}$,\newline 2h, $<5 \times 10^{-6}$ Torr &$0.82 \pm 0.06$& 66 \si{\mathrm{ions/spot}} &$440 \pm 40$&$4.0 \pm 0.6$\\

This work&2026&Electronic grade\newline bulk&FIB\newline 70, 35, 10, 5& 10-2000\newline \si{\mathrm{ions/spot}}& $\SI{1200}{\celsius}$,\newline 2h,\newline $\SI{1500}{\celsius}$,\newline 1h,\newline $1.5 \times 10^{-7}$ mbar & 6-33\% & 10 \si{\mathrm{ions/spot}} \newline for 70 keV & 25.2 & 1.15\\
\br
\end{tabular}
\end{adjustbox}
\end{table}

Despite the promising GeV properties, two key challenges for the application in quantum technology remain: i) the reliable fabrication of GeV CCs with desirable optical properties, and ii) the placement at well-defined positions inside nano-photonic structures.
The most common way to create color centers in diamond, especially those based on heavier elements, is a two-step process based on ion implantation to introduce the impurity ion into the diamond while also creating vacancies and a subsequent thermal annealing to activate the CCs and recover the lattice damage \cite{Redolfi2025-ju}.
Ion implantation is a mature and reliable method that has long been employed in semiconductor technology to fabricate solid-state devices. By adjusting the implantation energy, one can set the depth of the created defects, while the implanted ion fluence determines their density.
However, integration of CC into photonic structures requires more precise control: both the spatial placement of the defects (with precision of less than 100\,nm in all directions) and the number of implanted ions must be carefully tuned. Several strategies have been proposed, such as implantation through nano-collimated apertures \cite{Pezzagna2011} or resist-based patterned masks \cite{sangtawesin2014highly}.
Among these, focused ion beam (FIB) implantation provides unique advantages \cite{lesik2013maskless,tamura2014array,schroder2017scalable,ZhouNJP18,Pugliese2025-le}. First, the use of relatively low ion energies (below 100 \,keV) limits straggling and collision cascades, thus reducing the positional uncertainty of the defects created. Second, FIB systems can focus beams to spot sizes in the range of 10-30\,nm, which is on the order of, or even smaller than, the intrinsic ion straggling.
Third, ion sources that deliver picoampere currents, combined with pattern generators operating at dwell times of 10 to 100\,ns, enable implantation fluences approaching the single-ion regime. Finally, as a direct-write technique, FIB enables maskless implantation at user-defined locations, with excellent alignment ensured by high-precision positioning systems such as laser-interferometric stages.

The annealing step that typically follows ion implantation is required to form optically active CCs. The heating mobilizes vacancies and displaced carbon atoms, thereby partially repairing the damage caused by implantation \cite{Scattolo2025}. Furthermore, this allows the vacancies to migrate towards the implanted ions, forming the intended CCs. The temperature and duration of the annealing influence the formation yield of the CCs, that is, the number of optically active color centers created per implanted ion.
In addition, the annealing conditions affect the remaining type and amount of lattice damage in the vicinity of CCs \cite{YamPRB13}. This is particularly important as lattice damage can introduce noise that degrades the performance of nearby CCs.

Fabrication protocols for GeV centers in diamond have been extensively investigated in the literature to improve formation yield and establish reliable, efficient, and standardized processes for their successful integration into photonic structures \cite{Redolfi2025-ju}. Table \ref{table:summary} shows a summary of different reported fabrication methods and the achieved formation yield for GeV centers, including the relevant fabrication parameters.

In this work, we report an exceptionally high formation yield of negatively charged germanium vacancy centers achieved via optimized FIB implantation followed by high-temperature annealing. 
Creation of multiple arrays of Ge-implanted spots on the diamond substrate allows the study of the impact of implantation energies (in the range of 5\,keV to 70\,keV) and ion fluence (from 10 to 2000\, ions per spot) on the creation of GeV centers. Tuning the implantation fluence down to 10 ions per spot leads to the successful formation of single GeV in areas implanted at 35 and 70\,keV. Two methods were implemented to estimate the formation yield, with values ranging from 6\,\% to 32\,\% depending on the implantation parameters. Additionally, we report the successful creation of GeV centers by implanting at 5\,keV, thereby achieving a depth of only 5.5\,nm.

\section{Sample preparation and methods}

The experiments were performed on a commercially available 2x2x0.5\,mm$^{3}$ type IIa electronic-grade single-crystal (100)-oriented diamond substrate from ElementSix.

\subsection{Ion implantation}

The $^{74}$Ge implantation was performed with a commercially available Raith Velion system consisting of a perpendicular focused ion beam and a tilted scanning electron microscope (FIB-SEM) \cite{bauerdick2013multispecies}. The FIB is equipped with a liquid metal alloy ion source (LMAIS) \cite{bischoff2016liquid}, which is made of Gold, Germanium, and Silicon, referred to as AuGeSi source, and operates at a maximum acceleration voltage of 35\,kV.
A Wien filter allows selection of the desired positively charged ion species, and by choosing between single- and double-charged ions, it can cover a wider range of implantation energies, from 5 to 70\,keV. In the present work, germanium ions were implanted at four different energies (5, 10, 35, and 70\,keV) in separate areas of the sample. To identify the different regions, markers corresponding to the implantation energy were patterned into the diamond using Ge ions at 70\,keV.
Each of the four areas was implanted with an identical pattern consisting of nine spot arrays. Each array comprised of 100 ($5 \times 20$) spots implanted with Ge ions, along with an additional marker. Each spot was irradiated in the spot mode. In this mode, the beam position is fixed, and the ion-beam shutter is opened for the required dwell time to reach the desired ion fluence. The arrays were implanted with different fluences ranging from 10 to 2000 Ge ions per spot (see Figure \ref{fig:Fig_implantation}(a) in Appendix.)
To minimize contamination from neutrals originating from the AuGeSi ion source, all arrays were positioned at the edge of the FIB writing field ($200 \times 200$\,$\mu$m$^2$). The implantations were performed using a brand-new aperture to prevent beam deformation and improve the Wien filter's efficiency.
For each implantation energy, a different beam spot size was estimated using the “knife edge method” according to ISO/TR 19319:2013, resulting in a beam spot size of 20\,nm for 70 and 35\,keV, 60\,nm for 10\,keV, and 100\,nm for 5\,keV.
Monte Carlo simulations performed using the SRIM (Stopping and Range of Ions in Matter) software \cite{ziegler2010srim} estimated implantation depths ranging from 5.5 to 29.8\,nm, assuming a diamond density of 3.52\,g/cm$^3$, a displacement energy of 40\,eV, a surface binding energy of 7.41\,eV, a lattice binding energy of 7.62\,eV, and negligible channeling effects. The implantation conditions and expected implantation depth are summarized in Table \ref{table:implantation-parameter}.

\begin{table}
\label{table:implantation-parameter}
\caption{\label{label}Ion implantation specifications and expected depth distribution parameters from SRIM simulations.}
\begin{tabular}{@{}cccccc}
\br
Ion Energy&Acceleration&Ion Species&Beam spot&Projected range&Radial\\
(keV)&Voltage (kV)&&size (nm)&(nm)&straggling (nm)\\
\mr
70&35&Ge$^{++}$&20&29.8$\pm$7.0&6.7$\pm$3.6\\
35&17.5&Ge$^{++}$&20&17.5$\pm$4.0&4.2$\pm$2.2\\
10&5&Ge$^{++}$&60&7.8$\pm$1.6&2.1$\pm$1.1\\
5&5&Ge$^{+}$&100&5.5$\pm$1.0&2.1$\pm$1.1\\
\br
\end{tabular}
\end{table}

\subsection{Annealing}

After implantation, the sample was cleaned in a boiling triacid mixture (1:1:1 sulfuric, perchloric, and nitric acid) at 170$\si{\celsius}$ for 30 min. Subsequently, it was annealed in a two-step process in high vacuum (p $<$ 1.5 $\cdot$ 10$^{-7}$ mbar): first for 2 h at 1200$\si{\celsius}$, followed by 1 h at 1500°C (see Figure \ref{fig:Fig_implantation} (b) in the appendix). A similar process has been shown to yield SiV centers with good optical properties \cite{Lang_SiV} and was slightly modified for technical reasons. Afterward, the sample was cleaned again in the same triacid process as before, to remove any graphite formed on the surface during annealing and to obtain an oxygen-terminated surface \cite{Fuhrmann2024NVsurface}.

\subsection{Experimental setup}

\begin{figure}
    \centering
    \includegraphics[width=1\linewidth]{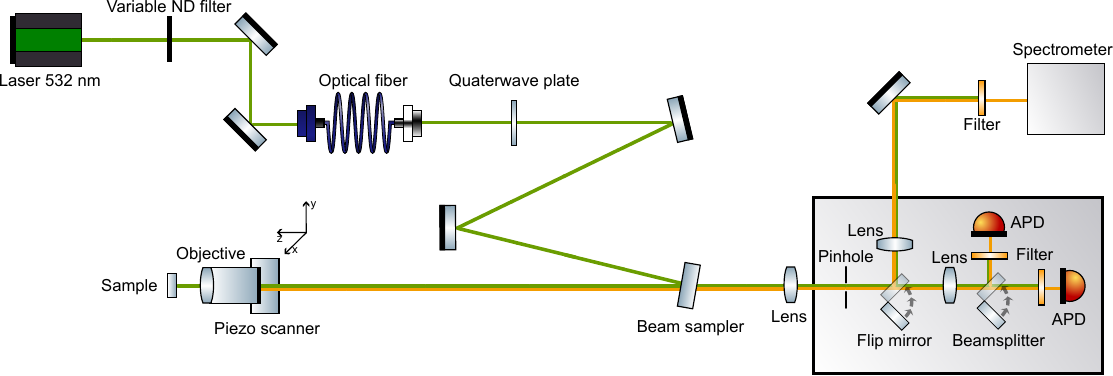}
    \caption{Schematic of the confocal setup. The excitation laser radiation is depicted in green, and the PL emission in orange.}
    \label{fig:Fig_Setup}
\end{figure}

The optical measurements were performed on a home-built confocal microscope. The optical scheme is shown in Figure\,\ref{fig:Fig_Setup}. GeV centers were excited off-resonantly with 532\,nm continuous-wave (CW) laser radiation. The laser light is coupled to the single-mode polarization-maintaining optical fiber in order to maintain the Gaussian profile of the excitation beam. After the fiber, the laser beam passes through the $\lambda/4$ plate to achieve circular polarization. The laser beam is reflected by the beam sampler and guided to the air objective (50x, NA = 0.95) mounted on the 3-axis piezo stage. The luminescent light is collected by the same objective and focused through a $15\,\mu m$-diameter pinhole. After passing through the pinhole, the fluorescent photons are detected by avalanche photodiodes (APDs). The 50:50 beamsplitter mounted on the motorized holder allows a fraction of the fluorescence light to be sent to the second APD for autocorrelation measurements. The 599/13 bandpass filters, corresponding to the ZPL emission of GeV, are installed in front of the APDs for fluorescence detection. All measurements were performed with a laser power of 0.8\,mW. In addition, fluorescent light can be directed to an optical spectrometer via a motorized flip mirror installed after the pinhole. The spectrometer was based on the Czerny-Turner monochromator and an actively cooled CCD camera. The photoluminescence spectra were recorded with a 300\,l/mm (blaze 500\,nm) diffraction grating. The residual laser light is blocked by a 550\,nm long-pass filter installed in front of the spectrometer. All confocal measurements were made using the modular software package Qudi \cite{QUDI17}.

\section{Results}

\subsection{Confocal Characterization}

\begin{figure}
    \centering
    \includegraphics[width=1\linewidth]{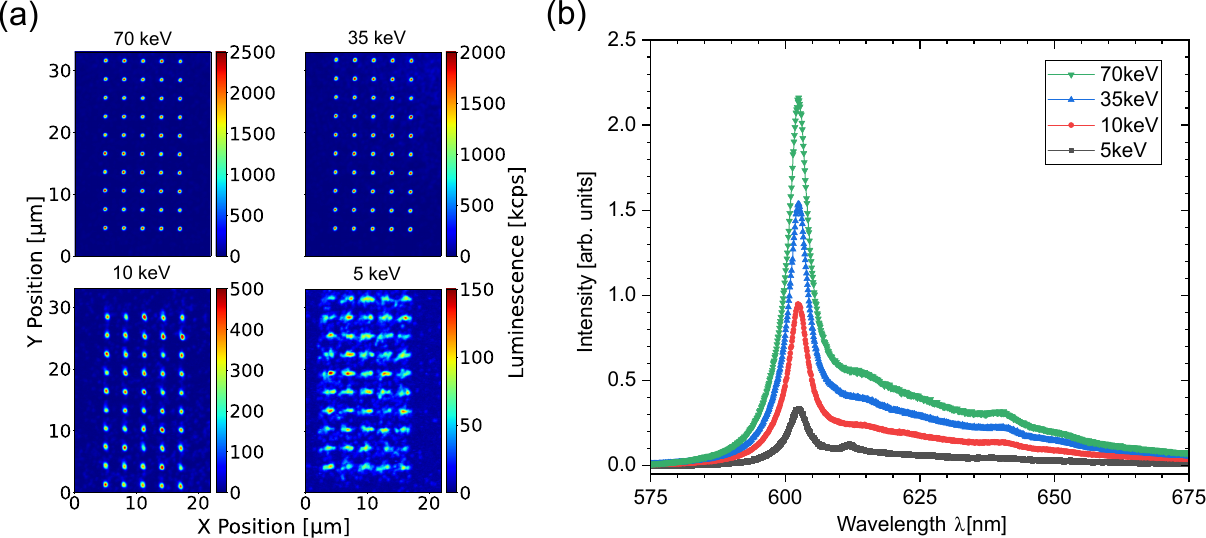}
    \caption{(a) PL images of implanted spot arrays of the 4 energies (5, 10, 35, 70\,keV). Each spot in the arrays was implanted with 1000 Ge ions. (b) Spectra taken in the corresponding regions all show the characteristic zero-phonon line at 602\, confirming the successful creation of Germanium vacancy centers.}
    \label{fig:Fig_PL3}
\end{figure}

To confirm the successful formation of GeV centers, confocal scanning was performed in the implanted areas. Figure \ref{fig:Fig_PL3} (a) shows the examples of recorded photoluminescence (PL) maps for areas implanted with 1000 ions per spot, but with different energies. For implantation energies of 70 and 35\,keV, clear, round-shaped spots are observed, indicating a well-focused ion beam. At lower implantation energies, the shapes deviate from the desired shape. In the 10\,keV area, spots show only small deviations from a round shape and appear slightly larger. In the 5\,keV region, the implantation spots are still recognizable but strongly elongated. This can be attributed to increased imperfections in ion-beam focusing. The main challenge arises from the focus optimization, which relies on secondary-electron images generated by the ion beam itself. The combination of lower-quality SE images and a low-energy ion beam makes it difficult to focus and achieve good implantation spots.
Furthermore, the implanted spot arrays in the 70 and 35\,keV areas are clearly identifiable across all fluences, including those implanted with only 10 ions per spot. In areas implanted at lower energies, the arrays can only be clearly identified starting at fluences of 50 ions per spot at 10\,keV and 200 ions per spot at 5\,keV. This indicates an energy dependence of the formation yield, with high values at 70 and 35\,keV.
To confirm that the emission from observed spots corresponds to GeV centers, PL spectra were acquired in all implantation regions. The examples of recorded spectra are shown in Figure \ref{fig:Fig_PL3} (b). All spectra show the characteristic shape with a peak at around 602\,nm corresponding to the zero-phonon line (ZPL) of GeV centers. This confirms the successful creation of germanium vacancy centers for all 4 implantation energies, including shallow GeV centers with an estimated depth of around 5.5\,nm (see Table \ref{table:implantation-parameter}).

\subsection{Characterization of Single Centers}

\begin{figure}
    \centering
    \includegraphics[width=1\linewidth]{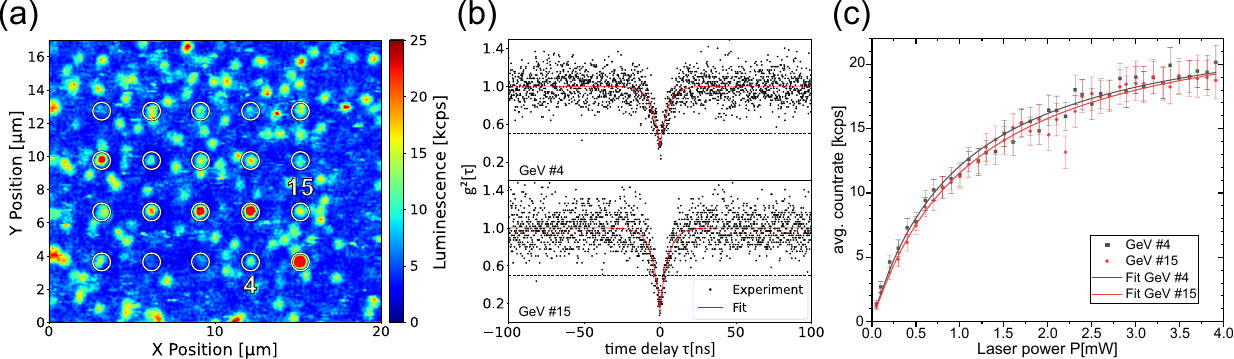}
    \caption{(a) PL map of the region implanted with 10 ions per spot and an implantation energy of 70keV. The circles indicate the implanted spots. (b) 2nd order autocorrelation curves of the implantation spots 4 and 15. A clear dip below 0.5 confirms the creation of single GeV centers.
    (c) Saturation curves acquired from the same GeV centers as in panel (b).}
    \label{fig:Fig_g22}
\end{figure}

To identify the successful creation of single GeV centers and, therefore, single-photon emitters (SPE), the low-fluence regions are further investigated. The second-order autocorrelation was measured at the 50 ions/spot array in the 10\,keV region and the 10 ions/spot arrays in the 35 and 70\,keV regions. The 5\,keV region was omitted because it was difficult to distinguish the implanted GeV centers from the background. For all 3 regions, a $5 \times 4$ array of implantation spots was evaluated.
An example of a PL map recorded in the 70\,keV region is shown in figure \ref{fig:Fig_g22} (a). The implantation pattern can clearly be identified. However, the area contains additional bright spots. The corresponding PL spectra reveal that the signal also arises from GeV. Those GeV centers presumably originate from an additional unintended implantation of neutral Ge ions that cannot completely be suppressed by the Wien filter in the FIB-SEM system.
The recorded normalized data is evaluated by fitting according to \cite{ZhouNJP18,schroder2017scalable}:
\begin{equation}
g^{2}(\tau) = 1 - A \cdot e^{- |\tau / \tau_{anti}| } + B \cdot e^{- |\tau / \tau_{bunch}| }
\end{equation}
where $\tau_{anti}$ and $\tau_{bunch}$ are the anti-bunching and bunching time constants, and A and B are the anti-bunching and bunching fit parameters. All of the fitting parameters depend on the internal relaxation times and the excitation intensity.

The depth of the dip can then be evaluated using $g^{2}(0) = 1 - 1/N$, where N is the number of emitters \cite{Tamura_2014g2,Meijer2005g2}. Therefore, if the dip is below 0.5, then the measured spot contains a single GeV center. Accordingly, this method can be used to identify spots with a low amount of spatially overlapping emitters. Nevertheless, this method is limited to a small number of emitters because the difference in dip depth decreases as the emitter count increases. Hence, the fewer emitters, the higher the precision of this method. Empirically, this method is therefore applicable for approximately 4 emitters.

Example of the experimental data for two single GeV centers in the 70\,keV region are shown in figure \ref{fig:Fig_g22} (b), both exhibit a $g^{2}(0)$ below 0.5. The evaluated array of the 70 and 35\,keV regions, each implanted with a total of 200 Ge ions, yielded 27 and 17 optically active GeV centers, respectively. The evaluated array in the 10\,keV region contained 51 germanium vacancy centers and was implanted with a total number of 1000 ions. This yields a first rough estimate of the formation yields of 13.5\,\% at 70\,keV, 8.5\,\% at 35\,keV, and 5.1\,\% at 10\,keV for the evaluated arrays.

Furthermore, the saturation behavior of the two single GeV centers shown in Figure \ref{fig:Fig_g22} (b) was investigated. The saturation curves were recorded by measuring the PL intensity I as a function of the laser excitation power. The PL intensity was determined by averaging the photon counts recorded over 30\,s and subtracting the background measured in the close vicinity. The results are shown in Figure \ref{fig:Fig_g22} (c) and exhibit clear saturation behavior. For the evaluation of the saturation, the data were fitted according to the following equation: 
\begin{equation}
I(P) = I_{sat} \cdot P/(P + P_{sat})
\end{equation}
where $I_{sat}$ is the countrate in saturation, $P_{sat}$ is the saturation power, and $P$ is the excitation power. This resulted in a saturation excitation power $P_{sat}$ of 1.12\,mW and 1.18\,mW for GeV \#4 and GeV \#15, as well as an optical saturation intensity $I_{sat}$ of 25.3\,kcps and 25.1\,kcps, respectively.

\subsection{Estimation of formation yield}

To study in detail the influence of implantation parameters on the formation yield, data taken from all implanted areas were evaluated. To increase the statistical significance of the estimated formation yield values, two methods have been implemented. The first method is based on comparing the photoluminescence signal recorded at an individual implanted spot with that from a single GeV center. This method is referred to as the photon yield method. The second method evaluates the integrated signal of a photoluminescence (PL) map of an implanted spot and compares it with that of a single GeV center. This method is referred to as the 2D PL map method.

\subsubsection{Photon Yield Method}

In this method, the formation yields are calculated as the ratio of the average photon count rate at the implanted spots to that of single-GeV centers. Due to the implantation beam size and ion straggling, one can expect the created GeV centers to be located within a volume significantly smaller than the confocal volume. Thus, one can assume that the emission from the brightest position in a spot contains contributions from all formed GeV centers.
For the photon yield method, the photon count rate of multiple implantation spots across all combinations of implantation parameters was recorded and averaged over 30 seconds. Additionally, the background for each spot is recorded independently at an adjacent position within $1\, \mu m$. The background is used to correct the measured photon count rate. To estimate the formation yield of an implantation spot, the number of GeV centers is calculated by dividing the photon count rate of the spot by the photon count rate of a single GeV center.
The formation yield estimates strongly depend on the precise determination of the countrate of a single GeV center. The investigation of the 26 identified single and 19 double GeV centers shows significant differences in their countrates, ranging from 8\,kcps to 19\,kcps for single centers. This can likely be attributed to differences in surface properties (e.g., roughness) or the local environment of the defects. This leads to significant uncertainty in the calculated average single-GeV count rate and, furthermore, explains the large error bars in the estimated formation yields. The resulting average single GeV photon count rate was calculated to be (12.5 $\pm$ 4.4)\,kcps.

\subsubsection{2D PL Map Method}

In contrast to the photon yield method, the 2D PL Map method is based on the analysis of the total emission from the implantation spot. For this, confocal images containing $5 \times 10$ implanted spots are recorded from each array. All images have an area of $22 \times 33\,$\textmu m$^2$ recorded with the resolution of $220 \times 330$ pixels$^2$ with constant laser power of 0.8\,mW. Thus, the images acquired from different implantation areas are directly comparable. The blob DoG (Difference of Gaussians) algorithm is used to define the implanted spots in the recorded confocal maps. Additionally, the retrieved list of coordinates is checked for incorrectly identified or unmarked positions. The 2D Gaussian fits are applied to the defined spots to determine their center positions. Afterward, the fluorescence from a $21 \times 21$-pixel$^2$ area centered on the identified maximum is summed. The resulting photon count is then background-corrected using the remaining unused pixels in the confocal image. The same procedure is applied to the previously identified single- and double-GeV centers. The obtained average values are used to calculate the number of GeV centers in all implanted spots. This method is feasible for calculating the formation yield when the implanted area exceeds the confocal volume. In addition, it allows individual background correction of each confocal image.

\begin{figure}
    \centering
    \includegraphics[width=1\linewidth]{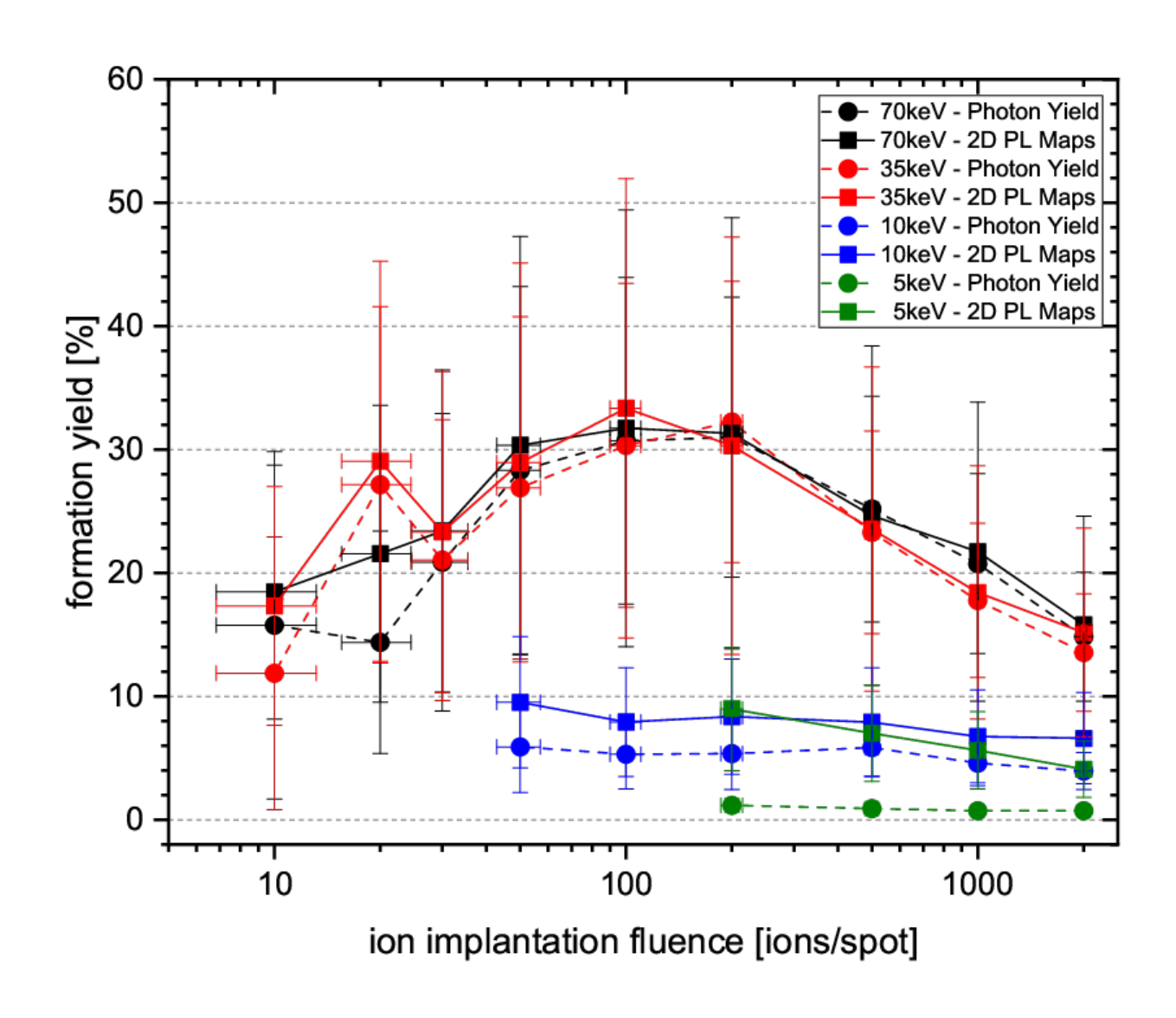}
    \caption{Dependence of the formation yield on the implantation fluence for all used ion energies. Round markers correspond to the values estimated by the photon yield method, whereas square markers correspond to the 2D PL method. Lines are drawn to guide the eyes.}
    \label{fig:Fig_yield}
\end{figure}

\subsection{Discussion}

The resulting GeV formation yields estimated by both methods are shown in Figure \ref{fig:Fig_yield} as a function of ion fluence for all implantation energies. Clearly, the formation yields show a significant difference between regions implanted with low (5 and 10\,keV) and high (35 and 70\,keV) energies.
For low implantation energies, the resulting formation yield depends on the method used. In particular, in the 5\, keV implanted region, the formation yield calculated using the photon-yield method is approximately $0.9\,\%$ and shows almost no dependence on the implantation fluence.
The 2D PL map method, on the other hand, estimates the formation yield to be $9.0\,\%$ at a fluence of 200 ions/spot, which gradually decreases to $4.1\,\%$ as the fluence increases.
This significant difference can be attributed to the shape of the implanted spots, see figure \ref{fig:Fig_PL3} (a). The implantation spots are large and elongated instead of small, round dots. The photon yield method, therefore, strongly underestimates the number of created GeV centers because only the brightest position of each implanted spot is taken into account. Thus, the PL signals from GeV centers outside of the confocal volume are not included in the estimation. The 2D PL map method, on the other hand, analyses the photon counts in a predefined area of pixels centered on the brightest pixel. Depending on the predefined area size, this method, unlike the photon yield method, can account for GeV centers that form farther from the brightest spot.
Therefore, the 2D PL map method accounts for the total emission within the predefined area, and contributions from all GeV centers can be accounted for in the recorded signal.
Note that accurately estimating the formation yield remains challenging because, even with the 2D PL map method, the results strongly depend on the ability to distinguish the intentionally formed GeV centers from the background. This makes it hard to find the correct center position and, therefore, to account for all GeV centers in the implantation spot.
The decrease in yield at higher fluences is therefore attributed to the difficulty and is assumed to be constant for the evaluated fluences. Henceforth, the yield is assumed to be the average of the estimated values, approximately 6.4$\,\%$.

In the region implanted with 10\,keV ions, confocal maps do not show significant distortion of the implanted spots (see figure \ref{fig:Fig_PL3} (a)). However, deviations from the expected round shape remain, especially compared with regions implanted with 35 and 70\,keV ions. This leads to a difference in the estimated formation yield between the two methods. In contrast to the 5\,keV region, both methods capture a similar dependence of the formation yield on the implantation fluence, with the yield nearly constant across all fluences. The averaged formation yield values differ slightly, i.e., approximately $5.2\,\%$ for the photon yield method and $7.8\,\%$ for the 2D PL map method, across all implantation fluences.
However, difficulties in yield estimation at low ion fluences persist, as noted for the 5\,keV implanted region. Thus, the 2D PL Map method is more representative at low implantation energies and yields formation yields of around $6.4\,\%$ and $7.8\,\%$ at implantation energies of 5 and 10\,keV, respectively, with no significant dependence on implantation fluence.

The formation yields estimated for high-energy-implanted areas differ noticeably from those obtained in low-energy regions. Due to the small and well-defined implantation spots, which can be resolved on PL maps (see figure \ref{fig:Fig_PL3} (a)), the yields can be estimated for the entire range of the implantation fluences. The formation yield shows a strong nonlinear dependence on the implantation fluence. For the lowest fluence (10\,ions$/$spot) the resulting formation yields are 14.6\,$\%$ and 17.1\,$\%$ for 35\,keV and 70\,keV, respectively. The formation yield increases with implantation fluence and reaches the highest values of 32.8\,$\%$ at 35\,keV and 31.4\,$\%$ at 70\,keV at a fluence of 100\,ions$/$spot. With further increases in the implantation fluence, the formation yields decrease to 14.4\,$\%$ and 15.3\,$\%$ at 35\,keV and 70\,keV, respectively. Both estimation methods provide similar results. This can be attributed to optimal focusing conditions during implantation, resulting in small, well-defined spots of in GeV centers.
In reported studies, either wide-beam implantation and high-temperature annealing \cite{Nieto_Hernandez2023-no, Hoy_Jensen2020-xv} or focused-ion-beam implantation and low-temperature annealing \cite{Pugliese2025-le, ZhouNJP18} was used, and all reported formation yields of below $1\,\%$ with the exception of \cite{Hoy_Jensen2020-xv}, that reported a formation yield of $6\,\%$ (see Table \ref{table:summary}).
Therefore, the significantly higher formation yields observed in the present work, as well as their fluence dependence, can be attributed to the interplay between high-temperature annealing and the specific conditions of FIB implantation.
For low implantation fluences, the formation yield is usually constant and decreases after reaching a certain threshold, e.g., as reported for the SiV center \cite{Lagomarsino_yield, schroder2017scalable}.
However, this work showed a strong dependence of the formation yield on implantation fluence, both at low and high fluences. Similar behavior was observed in \cite{Pezzagna_2010} and is attributed to the number of vacancies created in close proximity to the implanted ions. In the low-fluence regime, the number of vacancies created during implantation increases with increasing fluence. Thus, high-temperature annealing enables efficient trapping of vacancies by implanted ions, thereby increasing the formation yield. However, at a certain fluence, the lattice damage becomes detrimental, resulting in an impossible-to-anneal-out state that eventually turns the diamond into an amorphous state, which then converts to graphite upon thermal annealing \cite{Uzan-Saguy_graphite}. This results in a decrease in the formation yield at high ion fluences.   Ion implantation via a FIB can locally achieve high fluences, even in spot-irradiation mode, due to the extremely focused ion beam. For instance, 10 ions implanted with a 10-nm-diameter beam will result in a fluence of roughly $10^{13}$ cm$^{-2}$. 
Therefore, the maximum estimated formation yield can be reached relatively quickly with a well-focused ion beam, as seen in the 35 and 70\,keV implantation regions. Instead, the implantation with 5 and 10\,keV is carried out with much broader beams, i.e., lower ion density. This, together with the reduced number of vacancies created at lower kinetic energy, would explain the absence of a maximum in the formation yield.
The high-temperature annealing at 1500°C is highly effective in repairing damage induced by FIB implantation, which is especially important given the high local damage caused by the implantation. Furthermore, the increase in formation yield for all energies indicates that at the high annealing temperatures, the recombination of vacancies and Ge ions is highly effective.

\section{Conclusion}
In this work, the formation efficiency of GeV centers in diamond created by FIB implantation and high temperature annealing was studied. 
The diamond sample was implanted with germanium ions with energies of 5, 10, 35, and 70\,keV and different fluences. Subsequently, the sample was annealed at a maximum temperature of 1500\,°C to efficiently create GeV centers and recover the damaged diamond lattice.
The experimental results confirmed the successful creation of GeV centers across all implantation energies, thereby enabling the formation of shallow GeV centers at implantation energies as low as 5\,keV, corresponding to an implantation depth of only around 5.5\,nm.
Furthermore, confocal images of areas implanted at higher energies (35 and 70\,keV) suggest the possibility of creating GeV centers in small sample volumes, owing to the ion beam's small spot size and the ions' low lateral straggling. The investigation of the formation yield at low energies showed $~6\,\%$ at 5\,keV and $~8\,\%$ at 10\,keV across all ion fluences, representing an increase of 2-5x compared to previously reported values. The formation yield of the areas implanted at 35 and 70\,keV is even higher and shows a strong dependence on the implantation fluence. For low ion fluences, the yield is around $15\,\%$; whereas at higher fluences, it increases to a maximum of $~32\,\%$ for spots implanted with 100 or 200 ions, then decreases to $~16\,\%$ for spots implanted with 2000 ions, i.e. that is more then 10 times higher than previously reported. This is attributed to the strong interplay between the local implantation of FIB and the high temperature.

This method offers a highly efficient creation technique for GeV centers in small sample volumes with high lateral precision. This opens up the opportunity for precise incorporation into nanophotonic structures, such as photonic crystal cavities and pillars \cite{RedEPJQT25}.

\ack

This work was funded by the German Federal Ministry of Research, Technology, and Space (BMFTR) by the projects SPINNING (No. 13N16215), DE BRILL (No. 13N16207),  Quanten4KMU (No. 03ZU1110BB), QuantumHiFi (No. 16KIS1593), QR.N and CoGeQ, European Union's HORIZON Europe program via project SPINUS (No. 101135699), Deutsche Forschungsgemeinschaft (DFG) via projects No. 387073854, 546850640, and 491245864, and the joint DFG-Japan Science and Technology Agency (JST) project ASPIRE (No. 554644981).
E.S. and D.G. wish to acknowledge the EU HE project: Experimental production capabilities for quantum technologies in Europe, Qu-Pilot. Project ID: 101113983. E.M. and D.G. acknowledge financial support from PNRR MUR project PE0000023-NQSTI.

\appendix
\section{Additional Details Sample Preparation}

\begin{figure}
    \centering
    \includegraphics[width=1\linewidth]{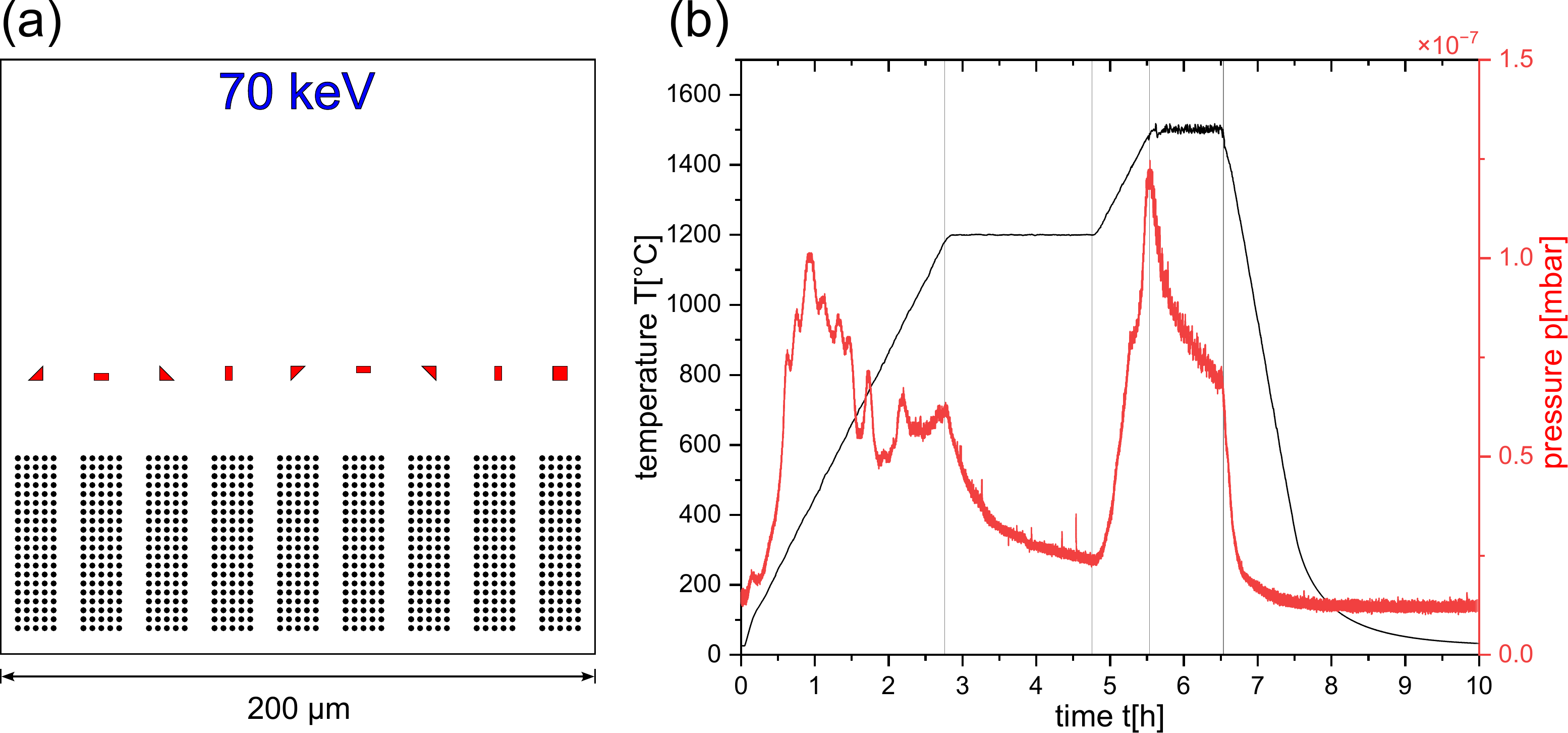}
    \caption{(a) Implantation map of the 70\,keV region with energy marker (blue), fluence markers (red), and the analyzed implanted spots (black). (b) Pressure and temperature measured during the annealing process.}
    \label{fig:Fig_implantation}
\end{figure}

Figure A1 (a) shows the implantation map used for all energies. Each area has dimensions of 200x200\,µm$^2$. In the top part, an energy marker is milled using Au ions. Furthermore, fluence markers were implanted using Ge ions to facilitate the identification of the spot arrays. The fluences used, starting from the left, are 2000, 1000, 500, 200, 100, 50, 30, 20, and 10\,ions/spot. All arrays were implanted using the spot irradiation mode of the FIB. Figure A1 (b) depicts the pressure and temperature measured during the annealing process. It can be seen that even for the annealing step at 1500°C, the pressure stays below 1.4e-7\,mbar.

\section*{References}

\bibliographystyle{ieeetr}
\bibliography{GeV_FIB}

\end{document}